\documentclass[onecolumn,10pt]{IEEEtran}

\usepackage{graphicx}
\usepackage{amsmath,amssymb}
\usepackage{cases}
\usepackage{color}
\usepackage{amsfonts}
\usepackage{indentfirst}
\usepackage{slashbox}
\usepackage{cite}
\usepackage{caption}
\usepackage{algorithm}
\usepackage{algpseudocode}
\usepackage{booktabs}
\usepackage{subfigure}
\usepackage{multirow}
\usepackage{amsthm}
\makeatletter
\newtheoremstyle{mythm}{3pt}{3pt}{}{16pt}{\bfseries}{:}{.5em}{}
\theoremstyle{mythm}
\newtheorem{theorem}{Theorem}
\newtheorem{example}{Example}
\newtheorem{definition}{Definition}
\newtheorem{remark}{Remark}

\newtheorem{construction}{Construction}
\begin{document}
\title{Constructions of Coded Caching Schemes with Flexible Memory Size
\author{Minquan Cheng, Jing Jiang, Qifa Yan, Xiaohu Tang, \IEEEmembership{Member,~IEEE}
}
\thanks{M. Cheng and J. Jiang are with Guangxi Key Lab of Multi-source Information Mining $\&$ Security, Guangxi Normal University,
Guilin 541004, China (e-mail: $\{$chengqinshi,jjiang2008$\}$@hotmail.com).}
\thanks{Q. Yan and X. Tang are with the Information Security and National Computing Grid Laboratory,
Southwest Jiaotong University, Chengdu, 610031, China (e-mail: qifa@my.swjtu.edu.cn, xhutang@swjtu.edu.cn).}
}
\date{}
\maketitle

\begin{abstract}
Coded caching scheme recently has become quite popular in the wireless network due to its effectively reducing the transmission amount (denote such an amount by $R$) during peak traffic times. However to realize a coded caching scheme, each file must be divided into $F$ packets which usually increases the computation complexity of a coded caching scheme.
So we prefer to construct a caching scheme that decreases the order of $F$ for practical implementations.

In this paper, we construct four classes of new schemes where two classes can significantly reduce the value of $F$ by increasing a little $R$ comparing with the well known scheme proposed by Maddah-Ali and Niesen, and $F$ in the other two classes grows sub-exponentially with $K$ by sacrificing more $R$. It is worth noting that a tradeoff between $R$ and $F$, which is a hot topic in the field of caching scheme, is proposed by our constructions. In addition, our constructions include all the results constructed by Yan et al., (IEEE Trans. Inf. Theory 63, 5821-5833, 2017) and some main results obtained by Shangguan et al., (arXiv preprint arXiv:1608.03989v1) as the special cases.
\end{abstract}

\begin{IEEEkeywords}
Coded caching scheme, placement delivery array, rate, packet number
\end{IEEEkeywords}

\section{Introduction}

Recently, the explosive increasing mobile services, especially applications such as video streaming, have imposed a tremendous pressure on the data transmission over the core network \cite{White}. As a result, during the peak-traffic times, the communication systems are usually congested. Coded caching scheme, which was proposed by Maddah-Ali and Niesen in \cite{AN}, can effectively reduce congestion during the peak-traffic times, and now is a hot topic in both industrial and academic fields (see \cite{MN1,RAN,YUTC,JCM ,KNMD ,Xiao2016tree ,GMDC,GGMG,JTLC,AN}, and references therein).

The benchmark work in \cite{AN} focused on the centralized caching system where a single server
containing $N$ files with the  same length connects to $K$ users over a shared link and each user has a cache memory of size $M$ files. A coded caching scheme consists of two phases: a placement phase during off-peak times and
a delivery phase during peak times. In the placement phase, the user caches are
populated. This phase does not depend on the user demands
which are assumed to be arbitrary. In delivery phase, each user requires a file from server. Then server sends a coded signal of at most $R$ files to the users such that various user demands are satisfied with the help the local caches. It is meaningful to minimize the load $R$ files in the delivery phase. Here $R$ is always called the rate. A coded caching scheme is called $F$-division scheme if each file is split into $F$ packets. If the packets of all files are directly cached in the placement phase, we call it uncoded placement. Otherwise we call it coded placement.
Through an elaborate uncoded placement and a coded delivery, the first determined scheme for an $F$-division $(K,M,N)$ coded caching system with $F={K\choose KM/N}$, when $\frac{KM}{N}$ is an integer, was proposed by Maddah-Ali and Niesen in \cite{AN}. Such a scheme is referred to as MN scheme in this paper.
By means of graph theory, reference \cite{WTP} showed that MN scheme has minimum rate under the constraint of uncoded cache placement when $K\leq N$. So far, many results have been obtained following MN schemes, for instances, \cite{AG,CYTJ,GR,STC,T,WLG,YTC} etc.

However it is easy to check that $F$ in MN scheme increases very quickly with the number of users $K$. This would become infeasible when $K$ becomes large. Furthermore, the complexity of a coded caching scheme increases with the parameter $F$.
Shanmugam et al. in \cite{SJTLD} first discussed such subpacketization problem for coded caching schemes. Very recently, Yan et al. in \cite{YCTC} characterized an $F$-division $(K,M,N)$ caching scheme by a simple array which is called $(K,F,Z,S)$ placement delivery array (PDA), where $\frac{M}{N}=\frac{Z}{F}$ and $R=\frac{S}{F}$. Then they proved that MN scheme is equivalent to a special PDA which is referred to as MN PDA. Furthermore, comparing with MN scheme they obtained two classes of coded caching schemes by means of constructing two infinite classes of PDAs such that $F$ reduces significantly by increasing the rate very little. Inspired by the concept of a PDA, we construct some coded caching schemes with lower level subpacketization were proposed in \cite{SZG,STD,TR,YCTC} from the view point of hypergraphs, resolvable combinatorial designs, Ruzsa-Szem\'{e}redi graphs, bipartite graphes respectively. For ease of exposition, the main previously known determined schemes are listed in Table \ref{tab-known}, which was summarised by Shangguang et al. in \cite{SZG}.
\begin{table}[H]
  \centering
  \caption{Summary of some known results \label{tab-known}}
  \normalsize{
  \begin{tabular}{|c|c|c|c|c|c|}
\hline
References &$K$  & $M/N$   & $R$   & $F$    \\ \hline
\cite{AN} &$K$&$\frac{z}{q}$, $ z=1,\ldots q-1$&$\frac{K(q-z)}{zK+q}$&${K\choose Kz/q}$\\ \hline \cite{YCTC,TR}&$(m+1)q$&$\frac{1}{q}$&$q-1$&$q^m$\\  \cline{2-5}
&$(m+1)q$&$\frac{q-1}{q}$&$\frac{1}{q-1}$& $(q-1)q^m$\\   \hline
\cite{SZG}&${m\choose t}q^t$&$1-(\frac{q-1}{q})^t$&$(q-1)^t$&$q^m$\\
 \cline{2-5}
&${m\choose t}q^t$&$1-\frac{1}{q^t}$&$\frac{1}{(q-1)^t}$&$(q-1)^t q^m$\\ \hline
\cite{STD}&$F^{\Omega(\frac{1}{\log\log F})}$  &$\frac{2}{3}+o(1)$ &$1$ &$F$\\   \hline
   \end{tabular}}
\end{table}

In this paper, we first propose two classes of schemes by directly constructing PDAs, i.e., Theorems \ref{th-general case 1} and \ref{th-special case 1}. Based on these two classes, another two classes of extended schemes, i.e., Theorems \ref{th-general case 2} and \ref{th-special case 2}, are obtained. We list these four classes of schemes in Table \ref{tab-main}.
\begin{table}[H]
  \centering
  \caption{Main constructions in this paper, where $z=1,2,\ldots,q-1$\label{tab-main}}
  \normalsize{
  \begin{tabular}{|c|c|c|c|c|c|}
\hline
Main results&$K$  & $M/N$   & $R$& $F$   \\ \hline
Theorem \ref{th-general case 1}&${m\choose t}q^t$  &$1-(\frac{q-z}{q})^t$& $\big(\frac{q-z}{\lfloor\frac{q-1}{q-z}\rfloor}\big)^t$&$\lfloor\frac{q-1}{q-z}\rfloor^t q^m$\\  \hline
Theorem \ref{th-special case 1}&$(m+1)q$  &$\frac{z}{q}$& $\frac{q-z}{\lfloor\frac{q-1}{q-z}\rfloor}$&$\lfloor\frac{q-1}{q-z}\rfloor q^{m}$\\   \hline
Theorem \ref{th-general case 2}&${m\choose t}\lfloor\frac{q-1}{q-z}\rfloor^t q^t$&$1-(\frac{q-z}{q})^t$&$(q-z)^t$&$q^m$\\  \hline
Theorem \ref{th-special case 2}&$(m\lfloor\frac{q-1}{q-z}\rfloor+1)q$&$\frac{z}{q}$&$q-z$&$q^m$\\  \hline
   \end{tabular}}
\end{table}
\noindent From Table \ref{tab-main}, it is not difficult to check that our schemes have more flexible $\frac{M}{N}$ than that of the results in \cite{STD,SZG,YCTC} and include all the results of \cite{SZG,YCTC} listed in Table \ref{tab-known} as special cases. Because of the flexible memory size, our new schemes have the following two important advantages.
\begin{itemize}
\item For the fixed $K$, $\frac{M}{N}$ we can obtain more different types of schemes directly. For example, given $K=405$ and $\frac{M}{N}=\frac{2}{3}$. Then from the results in \cite{SZG} and \cite{STD}, we have two types of schemes directly,
    while according to the results in Table \ref{tab-main}, the following $14$ schemes can be obtained.
\begin{small}
\begin{table}[H]
  \centering
  \caption{PDAs obtained by choosing different values $q$ and $z$ in Table \ref{tab-main}} \label{tab-flx2}
  \normalsize{
\begin{tabular}{|c|c|c|c|c|c|}
\hline
$q$&$z$&$m$                          &$R$&$\ln F$ \\ \hline
3  &2 &134&0.50&147.9072\\ \hline
3  &2 &67 &1   &73.6070\\ \hline
15 &10&26 &2.50&71.1025\\ \hline
9  &6 &22 &3   &48.3389\\ \hline
27 &18&14 &4.50&46.8349\\ \hline
15 &10&13 &5   &35.2047\\ \hline
45 &30&8  &7.50&31.1464\\ \hline
27  &18&7 &9   &23.0709\\ \hline
81 &54&4  &13.5&18.2709\\ \hline
45 &30&4  &15  &15.2266\\ \hline
135&90&2  &22.5&10.5037\\ \hline
81 &54&2  &27  &8.7889 \\ \hline
135&90&1  &45  &4.9053 \\ \hline
\end{tabular}}
\end{table}
\end{small}
\item For many values of $\frac{M}{N}$, the schemes in Theorem \ref{th-general case 1} have both smaller $F$ and $R$ than those of the schemes generated by memory sharing based on \cite{SZG} listed in Table \ref{tab-known}, and the schemes in
    Theorems \ref{th-special case 1} have both smaller $F$ and $R$ than those of the schemes generated by memory sharing based on \cite{YCTC} listed in Table \ref{tab-known}. The detailed discusses are proposed in Subsections \ref{sub-per1.2} and \ref{sub-per1}.
\end{itemize}

The rest of this paper is organized as follows. Section \ref{sec_prob} introduces caching system model, placement delivery array and main result in this paper. In Section \ref{sec-performance}, the detailed performance analyses of Theorems \ref{th-general case 1} and \ref{th-special case 1} are proposed. In \ref{se-generalized}, we first propose the constructions for Theorems \ref{th-general case 1} and \ref{th-special case 1} respectively. Based on these two constructions, two extended constructions are proposed for Theorem \ref{th-general case 2} and \ref{th-special case 2}. Finally conclusion is drawn in Section \ref{conclusion}.

\section{System Model and Placement Delivery Array}\label{sec_prob}
In this paper, we use bold capital letter,  bold lower case letter and curlicue letter to denote array, vector and set respectively. We use $[a,b]=\{a,a+1,\ldots,b\}$ and $[a, b)=\{a,a+1,\ldots,b-1\}$ for intervals of integers for any integers $a$ and $b$ with $a\leq b$.

\subsection{System model}
Denote $N$ files by $\mathcal{W}=\{W_1,W_2,\cdots,W_N\}$ and $K$ users by $\mathcal{K}=\{1,2,\cdots,K\}$.
An $F$-division $(K,M,N)$ coded caching scheme proposed by Maddah-Ali and Niesen in \cite{AN} operates in two separated phases:
\begin{enumerate}
  \item \textbf{Placement Phase:} Each file is sub-divided into $F$ equal packets\footnote{Memory sharing technique may lead to non equally divided packets \cite{AN}, in this paper, we will not discuss this case.}, i.e., $W_{i}=\{W_{i,j}:j=1,2,\cdots,F\}$. Each user is accessible to the files set $\mathcal{W}$. Denote $\mathcal{Z}_k$ the packets subset of $\mathcal{W}$ cached by user $k$.
  \item \textbf{Delivery Phase:} Each user requests one file from $\mathcal{W}$ randomly. Denote the request by ${\bf d}=(d_1,d_2,\cdots,d_K)$, i.e., user $k$ requests file $W_{d_k}$, where $k\in \mathcal{K}, d_k\in\{1,2,\cdots,N\}$. Once the server receives the request ${\bf d}$, it broadcasts XOR of packets with size of at most $R_{{\bf d}}F$ to users such that each user is able to recover its requested file.

\end{enumerate}

Denote the maximum transmission amount among all the request during the delivery phase by $R$, i.e.,
\begin{align*}
R=\sup_{\tiny{\mbox{$\begin{array}{c}
                 \mathbf{d}=(d_1,\cdots,d_{K}) \\
                 d_k\in[1,N],\forall k\in[1,K]
               \end{array}
$}}} \left\{R_{\mathbf{d}}\right\}.
\end{align*}
$R$ is always called the rate of a coded caching scheme.
%

\subsection{Placement delivery array}
In this paper we focus on the caching system in the above subsection. Yan et al., in \cite{YCTC} proposed an interesting and simple combinatorial structure, called placement delivery array, which can characterize the placement phase and delivery phase simultaneously.
\begin{definition}(\cite{YCTC})
For  positive integers $K,F, Z$ and $S$, an $F\times K$ array  $\mathbf{P}=(p_{j,k})$, $1\leq j\leq F, 1\leq k\leq K$, composed of a specific symbol $``*"$  and $S$ positive integers
$1,2,\cdots, S$, is called a $(K,F,Z,S)$ placement delivery array (PDA) if it satisfies the following conditions:
\begin{enumerate}
  \item [C$1$.] The symbol $``*"$ appears $Z$ times in each column;
  \item [C2.] Each integer occurs at least once in the array;
  \item [C$3$.] For any two distinct entries $p_{j_1,k_1}$ and $p_{j_2,k_2}$,    $p_{j_1,k_1}=p_{j_2,k_2}=s$ is an integer  only if
  \begin{enumerate}
     \item [a.] $j_1\ne j_2$, $k_1\ne k_2$, i.e., they lie in distinct rows and distinct columns; and
     \item [b.] $p_{j_1,k_2}=p_{j_2,k_1}=*$, i.e., the corresponding $2\times 2$  subarray formed by rows $j_1,j_2$ and columns $k_1,k_2$ must be of the following form
  \begin{eqnarray*}
    \left(\begin{array}{cc}
      s & *\\
      * & s
    \end{array}\right)~\textrm{or}~
    \left(\begin{array}{cc}
      * & s\\
      s & *
    \end{array}\right).
  \end{eqnarray*}
   \end{enumerate}
\end{enumerate}
\end{definition}
\begin{theorem}(\cite{YCTC})
\label{th-Fundamental}Using Algorithm \ref{alg:PDA}, an $F$-division caching scheme for a $(K,M,N)$ caching system can be realized by a $(K,F,Z,S)$ PDA  with $\frac{M}{N}=\frac{Z}{F}$. Each user can decode his requested file correctly for any request ${\bf d}$ at the rate $R=\frac{S}{F}$.
\end{theorem}
\begin{algorithm}[htb]
\caption{caching scheme based on PDA in \cite{YCTC}}\label{alg:PDA}
\begin{algorithmic}[1]
\Procedure {Placement}{$\mathbf{P}$, $\mathcal{W}$}
\State Split each file $W_i\in\mathcal{W}$ into $F$ packets, i.e., $W_{i}=\{W_{i,j}\ |\ j=1,2,\cdots,F\}$.
\For{$k\in \mathcal{K}$}
\State $\mathcal{Z}_k\leftarrow\{W_{i,j}\ |\ p_{j,k}=*, \forall~i=1,2,\cdots,N\}$
\EndFor
\EndProcedure
\Procedure{Delivery}{$\mathbf{P}, \mathcal{W},{\bf d}$}
\For{$s=1,2,\cdots,S$}
\State  Server sends $\bigoplus_{p_{j,k}=s,1\leq j\leq F,1\leq k\leq K}W_{d_{k},j}$.
\EndFor
\EndProcedure
\end{algorithmic}
\end{algorithm}

\begin{example}\rm
\label{E-pda}
It is easy to verify that the following array is a $(4,6,3,4)$ PDA:
\begin{eqnarray*}
\mathbf{P}_{6\times 4}=\left(\begin{array}{cccc}
*&*&1&2\\
*&1&*&3\\
*&2&3&*\\
1&*&*&4\\
2&*&4&*\\
3&4&*&*
\end{array}\right).
\end{eqnarray*}
Using Algorithm \ref{alg:PDA}, one can obtain a $6$-division $(4,3,6)$ coded caching scheme in the following way.
\begin{itemize}
   \item \textbf{Placement Phase}: From Line 2 we have $W_i=\{W_{i,1},W_{i,2},W_{i,3},W_{i,4},W_{i,5},W_{i,6}\}$ , $i\in [1,6]$. Then by Lines 3-5, the contents in each user are
       \begin{align*}
       \mathcal{Z}_1=\left\{W_{i,1},W_{i,2},W_{i,3}:i\in[1,6]\right\}\ \ \ \ \ \ \
       \mathcal{Z}_2=\left\{W_{i,1},W_{i,4},W_{i,5}:i\in[1,6]\right\} \\
       \mathcal{Z}_3=\left\{W_{i,2},W_{i,4},W_{i,6}:i\in[1,6]\right\}\ \ \ \ \ \ \
       \mathcal{Z}_4=\left\{W_{i,3},W_{i,5},W_{i,6}:i\in[1,6]\right\}
       \end{align*}
   \item \textbf{Delivery Phase}: Assume that the request vector is $\mathbf{d}=(1,2,3,4,5,6)$. Table \ref{table1} shows the transmitting process by Lines 8-10.
   \begin{table}[!htp]

  \normalsize{
  \begin{tabular}{|c|c|}
\hline
   Time Slot& Transmitted Signnal  \\
\hline
   $1$&$W_{1,4}\oplus W_{2,2}\oplus W_{3,1}$\\ \hline
   $2$&$W_{1,5}\oplus W_{2,3}\oplus W_{4,1}$\\ \hline
  $3$& $W_{1,6}\oplus W_{3,3}\oplus W_{4,2}$\\ \hline
   $4$& $W_{2,6}\oplus W_{3,5}\oplus W_{4,4}$\\ \hline
  \end{tabular}}\centering
  \caption{Delivery steps in Example \ref{E-pda} }\label{table1}
\end{table}
\end{itemize}
\end{example}

From Theorem \ref{th-Fundamental}, an $F$-division $(K,M,N)$ coded caching scheme can be obtained by constructing an appropriate PDA. In this paper,
we only consider the caching schemes with $K\leq N$.
\subsection{Main Results}
In this paper, we directly construct the following two classes of PDAs.
\begin{theorem}
\label{th-general case 1}
For any positive integers $q$, $z$, $m$ and $t$ with $q\geq2$, $z<q$ and $t<m$, the array $\mathbf{P}$ given in Construction \ref{con-general-1} is an 
$({m\choose t}q^t$, $\lfloor\frac{q-1}{q-z}\rfloor^t q^m$, $\lfloor\frac{q-1}{q-z}\rfloor^t(q^m-q^{m-t}(q-z)^t)$, $(q-z)^tq^{m})$ PDA with $\frac{M}{N}=1-(\frac{q-z}{q})^t$ and rate $R=(q-z)^t/\lfloor\frac{q-1}{q-z}\rfloor^t$.
\end{theorem}

\begin{theorem}\label{th-special case 1}
For any positive integers $q$, $z$, $m$  with $q\geq2$ and $z<q$, the array $\mathbf{H}$ generated by Construction \ref{con-special-1} is an $((m+1)q,\lfloor\frac{q-1}{q-z}\rfloor q^{m},z\lfloor\frac{q-1}{q-z}\rfloor q^{m-1},(q-z)q^{m})$ PDA with $\frac{M}{N}=\frac{z}{q}$ and rate $R=(q-z)/\lfloor\frac{q-1}{q-z}\rfloor$.
\end{theorem}
Based on the above two results, the following extended results can be obtained.
\begin{theorem}
\label{th-general case 2}
For any positive integers $q$, $z$, $m$ and $t$ with $q\geq2$, $z<q$ and $t<m$, there exists an $({m\choose t}q^t\lfloor\frac{q-1}{q-z}\rfloor^t$, $q^m$, $q^m-(q-z)^t q^{m-t}$, $(q-z)^tq^{m})$ PDA with $\frac{M}{N}=1-(\frac{q-z}{q})^t$ and rate $R=(q-z)^t$.
\end{theorem}

\begin{theorem}\label{th-special case 2}
For any positive integers $q$, $z$, $m$ with $q\geq2$ and $z<q$, the array $(\mathbf{P},\mathbf{C})$ generated by Construction \ref{con-special-2} is an 
$((m\lfloor\frac{q-1}{q-z}\rfloor+1)q,q^{m},zq^{m-1},(q-z)q^{m})$ PDA with $\frac{M}{N}=\frac{z}{q}$ and rate $R=q-z$.
\end{theorem}
\section{Performance Analyses}
\label{sec-performance}
\subsection{Performance analyses of Theorem \ref{th-general case 1}}
\label{sub-per1.2}
First we claim that $F$ of the PDA in Theorem \ref{th-general case 1} grows
sub-exponentially with $K$ when $t\geq 2$. Let $q$, $z$ and $t$ be fixed and let $m$ approximate infinity, it holds that $R$ and $\frac{M}{N}$ are constants independent of $K$ for all PDAs in Theorem \ref{th-general case 1}. Similar to the discussion in \cite{SZG}, by means of  the inequality $(m/t)^t<{m\choose t}<(em/t)^t$, we can estimate the value of $m$ by $K$, i.e.,
\begin{eqnarray}
\label{eq-ex-G-m}
\frac{tK^{\frac{1}{t}}}{eq}<m<\frac{tK^{\frac{1}{t}}}{q}.
\end{eqnarray}
So we have $F=\mathcal{O}(\lfloor\frac{q-1}{q_-z}\rfloor^t q^{\frac{tK^{1/t}}{q}})$. This implies that $F$ grows sub-exponentially with $K$ if $t\geq 2$.

Secondly let us consider the performance in Theorem \ref{th-general case 1} comparing the result in \cite{SZG} listed in Table \ref{tab-known}. When $\frac{M}{N}=1-(\frac{q-1}{q})^t$, $1-\frac{1}{q^t}$, from Theorem \ref{th-general case 1} we have
\begin{itemize}
\item a $({m\choose t}q^t,q^m,q^m-q^{m-t}(q-1)^t,(q-1)^tq^{m})$ PDA with $\frac{M}{N}=1-(\frac{q-1}{q})^t$ and rate $R=(q-1)^t$         and
\item a $({m\choose t}q^t$, $(q-1)^t q^m,(q-1)^t(q^m-q^{m-t}),q^{m})$ PDA with $\frac{M}{N}=1-\frac{1}{q^t}$ and rate $R=\frac{1}{(q-1)^t}$.
\end{itemize}
Clearly these two PDAs are exactly the results in \cite{SZG} listed in Table \ref{tab-known}. So we only need to consider the positive real number $1-(\frac{q-1}{q})^t <\frac{M}{N}<1-\frac{1}{q^t}$. Let $K={m\choose t}q^t$. Based on the schemes in  \cite{SZG}, we can obtain the schemes with $K$ users and memory ratio $\frac{M}{N}$ by memory sharing. Now let us introduce memory sharing method first. Assume that there exist $r$ schemes with $(\frac{M_i}{N},R_i,F_i)$, $i=1,2,\ldots,r$. \cite{AN} showed that
a scheme with
\begin{eqnarray}
\label{eq-M-MNRF}
\begin{split}
\frac{M}{N}&=\lambda_1 \frac{M_1}{N}+\lambda_2 \frac{M_1}{N}+\ldots+\lambda_{r} \frac{M_{r}}{N}\\
R_{M}&=\lambda_1 R_1+\lambda_2 R_2+\ldots+\lambda_{r} R_{r}\\
F_{M}&=F_1+F_2+\ldots+F_{r}
\end{split}
\end{eqnarray}
can be obtained where $M_1\leq M\leq M_{r}$, $0< \lambda_i\leq 1$ and $\sum_{i=1}^{r}\lambda_i=1$. From the results in \cite{SZG} of Table \ref{tab-known}, there exist two schemes with
\begin{eqnarray}
\label{eq-SZG}
\left(\frac{M_{1}}{N},R_{1},F_{1}\right)=\left(1-(\frac{q-1}{q})^t, (q-1)^t, q^m\right) \ \ \hbox{and}\ \ \left(\frac{M_{2}}{N},R_{2},F_{2}\right)=\left(1-\frac{1}{q^t}, \frac{1}{(q-1)^t},(q-1)^t q^m\right).
\end{eqnarray}
So the scheme in \eqref{eq-M-MNRF} based on the schemes in \eqref{eq-SZG} has the parameters
\begin{eqnarray}
\label{eq-MN-M-R}
\begin{split}
\frac{M}{N}&=\lambda (1-(\frac{q-1}{q})^t)+(1-\lambda)\left(1-\frac{1}{q^t}\right)\\
R_{M-S}&=\lambda (q-1)^t+(1-\lambda) \frac{1}{(q-1)^t} \\
F_{M-S}&=(q-1)^t q^m+q^m
\end{split}
\end{eqnarray}
for some positive real number $0<\lambda<1$. From Theorem \ref{th-general case 1}, there exist $q-1$ schemes with
\begin{eqnarray}
\label{eq-Gen-1}
\left(\frac{M_{z}}{N},R_{z},F_{z}\right)=\left(1-(\frac{q-z}{q})^t,\big(\frac{q-z}{\lfloor\frac{q-1}{q-z}\rfloor}\big)^t,
\lfloor\frac{q-1}{q-z}\rfloor^t q^m\right),\ \ z=1,2,\ldots,q-1.
\end{eqnarray}
If there exists $z$, $1\leq z\leq q-1$, satisfying $\frac{M}{N}=1-(\frac{q-z}{q})^t$, by \eqref{eq-MN-M-R} and \eqref{eq-Gen-1} we have
\begin{eqnarray}
\label{eq-comR-S-C1-1}
\begin{split}
\frac{R_{z}}{R_{M-S}}&=\frac{\big(\frac{q-z}{\lfloor\frac{q-1}{q-z}\rfloor}\big)^t}{\lambda (q-1)^t+(1-\lambda) \frac{1}{(q-1)^t}}=\frac{1}{\left(\lambda (\frac{q-1}{q-z})^t+\frac{1-\lambda}{(q-1)^t(q-z)^t}\right)
\lfloor\frac{q-1}{q-z}\rfloor^t}\\
&<\frac{1}{\lambda (\frac{q-1}{q-z})^t\lfloor\frac{q-1}{q-z}\rfloor^t}<\frac{1}{\lambda \lfloor\frac{q-1}{q-z}\rfloor^{2t}}
\end{split}
\end{eqnarray}
and
\begin{eqnarray}
\label{eq-comF-S-C1-1}
\begin{split}
\frac{F_{z}}{F_{M-S}}&=\frac{\lfloor\frac{q-1}{q-z}\rfloor^t q^m}{(q-1)^t q^m+q^m}=\frac{\lfloor\frac{q-1}{q-z}\rfloor^t }{(q-1)^t+1}=\frac{1}{\left(\frac{q-1}{\lfloor\frac{q-1}{q-z}\rfloor}\right)^t+\frac{1}{\lfloor\frac{q-1}{q-z}\rfloor^t}}\\
&<\frac{1}{(q-z)^t+\frac{1}{\lfloor\frac{q-1}{q-z}\rfloor^t}}<\frac{1}{(q-z)^t}.
\end{split}
\end{eqnarray}
\begin{remark}
\label{re-1}
From \eqref{eq-comR-S-C1-1} and \eqref{eq-comF-S-C1-1}, when $\lambda \lfloor\frac{q-1}{q-z}\rfloor^{2t}>1$, $R_z$ and $F_z$ are at least $\frac{1}{\lambda \lfloor\frac{q-1}{q-z}\rfloor^{2t}}$ and $\frac{1}{(q-z)^t}$ times smaller than $R_{M-S}$ and $F_{M-S}$ respectively.
\end{remark}
\begin{example}
\label{ex-com1}
When $t=3$, $q=20$ and $\lambda=0.1$, the following table can be obtained by \eqref{eq-comR-S-C1-1} and \eqref{eq-comF-S-C1-1}.
\begin{table}[H]
  \centering
  \caption{Comparisons of the performances between schemes from \cite{SZG} and Theorem \ref{th-general case 1}} \label{tab_compare th3,5}
  \normalsize{
  \begin{tabular}{|c|c|c|}
\hline
$z$&  $\frac{R_{z}}{R_{M-S}}<$    &$\frac{F_{z}}{F_{M-S}}$   \\ \hline
11 &0.15625&0.00137174211248285\\ \hline
12 &0.15625&0.001953125\\ \hline
13 &0.15625&0.00291545189504373\\
14 &0.0137174211248285 & 0.00462962962962963\\ \hline
15 &0.0137174211248285 & 0.008\\ \hline
16 &0.00244140625      &0.015625\\ \hline
17 &0.000214334705075446 &        0.037037037037037\\ \hline
18 &1.88167642315892e-05 & 0.125\\ \hline
\end{tabular}}
\end{table}

\end{example}

If there is no integer $z$ satisfying $\frac{M}{N}=1-(\frac{q-z}{q})^t$, a scheme in \eqref{eq-M-MNRF} based on the schemes in \eqref{eq-Gen-1} has the parameters
\begin{eqnarray}
\label{eq-MN-M-Gen-1}
\begin{split}
\frac{M}{N}&=\lambda'\left(1-(\frac{q-z}{q})^t\right)+(1-\lambda')\left(1-(\frac{q-z-1}{q})^t\right)\\
R_{M-C_1}&=\lambda' \big(\frac{q-z}{\lfloor\frac{q-1}{q-z}\rfloor}\big)^t+(1-\lambda') \big(\frac{q-z-1}{\lfloor\frac{q-1}{q-z-1}\rfloor}\big)^t< \big(\frac{q-z}{\lfloor\frac{q-1}{q-z}\rfloor}\big)^t\\
F_{M-C_1}&=\lfloor\frac{q-1}{q-z}\rfloor^t q^m+\lfloor\frac{q-1}{q-z-1}\rfloor^t q^m
\end{split}
\end{eqnarray}
for some positive integer $z$ with $1-(\frac{q-z}{q})^t<\frac{M}{N}<1-(\frac{q-z-1}{q})^t$ and a real number $0<\lambda'< 1$. Here the last item in the second formula of \eqref{eq-MN-M-Gen-1} holds since
$\big(\frac{q-z}{\lfloor\frac{q-1}{q-z}\rfloor}\big)^x>\big(\frac{q-z-1}{\lfloor\frac{q-1}{q-z-1}\rfloor}\big)^x$ always holds for any positive integer $x$. Similar to \eqref{eq-comR-S-C1-1} and \eqref{eq-comF-S-C1-1}, we have
\begin{eqnarray}
\label{eq-comR-S-C1-2}
\frac{R_{M-C_1}}{R_{M-S}}&<\frac{\big(\frac{q-z}{\lfloor\frac{q-1}{q-z}\rfloor}\big)^t}{\lambda (q-1)^t+(1-\lambda) \frac{1}{(q-1)^t}}<\frac{1}{\lambda \lfloor\frac{q-1}{q-z}\rfloor^{2t}}
\end{eqnarray}
and
\begin{eqnarray}
\label{eq-comF-S-C1-2}
\begin{split}
\frac{F_{M-C_1}}{F_{M-S}}&=\frac{\lfloor\frac{q-1}{q-z}\rfloor^t q^m+\lfloor\frac{q-1}{q-z-1}\rfloor^t q^m}{(q-1)^t q^m+q^m}=\frac{\lfloor\frac{q-1}{q-z}\rfloor^t +\lfloor\frac{q-1}{q-z-1}\rfloor^t }{(q-1)^t+1}=\frac{\lfloor\frac{q-1}{q-z}\rfloor^t}{(q-1)^t+1} +\frac{\lfloor\frac{q-1}{q-z-1}\rfloor^t }{(q-1)^t+1}\\
&<\frac{1}{(q-z)^t}+\frac{1}{(q-z-1)^t}.
\end{split}
\end{eqnarray}
Similar to the Remark \ref{re-1}, we have that when $\lambda \lfloor\frac{q-1}{q-z}\rfloor^{2t}>1$, $R_{M-C_1}$ and $F_{M-C_1}$ are at least $\frac{1}{\lambda \lfloor\frac{q-1}{q-z}\rfloor^{2t}}$ and $\frac{1}{(q-z)^t}+\frac{1}{(q-z-1)^t}$ times smaller than $R_{M-S}$ and $F_{M-S}$ respectively.
\subsection{Performance analyses of Theorem \ref{th-special case 1}}
\label{sub-per1}
For any positive integers $m$ and $q$, assume that $K=(m+1)q$. Similar to the discussion in Subsection \ref{sub-per1.2}, we also first consider the case $\frac{M}{N}=\frac{1}{q}$ and $\frac{q-1}{q}$. From Theorem \ref{th-special case 1} we have
\begin{itemize}
\item an $((m+1)q,q^{m},q^{m-1},(q-1)q^{m})$ PDA with $\frac{M}{N}=\frac{1}{q}$ and $R=q-1$, and
\item an $((m+1)q,(q-1)q^{m},(q-1)^2q^{m-1},q^{m})$ PDA with $\frac{M}{N}=1-\frac{1}{q}$ and rate $R=\frac{1}{q-1}$.
\end{itemize}
Clearly these two PDAs are exactly the results in \cite{YCTC} listed in Table \ref{tab-known}. It is worth to note that the authors in \cite{YCTC} showed that comparing with MN PDA, the packet number $F$ of these two PDAs in Theorem \ref{th-special case 1} reduces significantly while the rate $R$ increases little. Now let us consider the case $\frac{1}{q}<\frac{M}{N}<\frac{q-1}{q}$.

From the results in \cite{YCTC} of Table \ref{tab-known}, there exist two schemes with
\begin{eqnarray}
\label{eq-YCTC}
\left(\frac{M_{1}}{N},R_{1},F_{1}\right)=\left(\frac{1}{q}, q-1, q^m\right) \ \ \hbox{and}\ \ \left(\frac{M_{2}}{N},R_{2},F_{2}\right)=\left(1-\frac{1}{q}, \frac{1}{q-1},(q-1)q^m\right).
\end{eqnarray}
So the scheme in \eqref{eq-M-MNRF} based on the schemes in \eqref{eq-SZG} has the parameters
\begin{eqnarray}
\label{eq-MN-M-R1}
\begin{split}
\frac{M}{N}&=\lambda \frac{1}{q}+(1-\lambda)\left(1-\frac{1}{q}\right)\\
R_{M-Y}&=\lambda (q-1)+(1-\lambda) \frac{1}{q-1} \\
F_{M-Y}&=(q-1)q^m+q^m
\end{split}
\end{eqnarray}
for some positive real number $0<\lambda<1$. From Theorem \ref{th-general case 1}, there exist $q-1$ schemes with
\begin{eqnarray}
\label{eq-Sep-1}
\left(\frac{M_{z}}{N},R_{z},F_{z}\right)=\left(\frac{z}{q},\frac{q-z}{\lfloor\frac{q-1}{q-z}\rfloor},
\lfloor\frac{q-1}{q-z}\rfloor q^m\right),\ \ z=1,2,\ldots,q-1.
\end{eqnarray}
If there exists $z$, $1\leq z\leq q-1$, satisfying $\frac{M}{N}=\frac{z}{q}$, by \eqref{eq-MN-M-R1} and \eqref{eq-Sep-1} we have
\begin{eqnarray}
\label{eq-comR-Y-C1-1}
\begin{split}
\frac{R_{z}}{R_{M-Y}}&=\frac{\frac{q-z}{\lfloor\frac{q-1}{q-z}\rfloor}}{\lambda (q-1)+(1-\lambda) \frac{1}{(q-1)}}=\frac{1}{\left(\frac{\lambda(q-1)}{q-z}+\frac{1-\lambda}{(q-1)(q-z)}\right)
\lfloor\frac{q-1}{q-z}\rfloor}\\
&<\frac{1}{\lambda (\frac{q-1}{q-z})\lfloor\frac{q-1}{q-z}\rfloor}<\frac{1}{\lambda \lfloor\frac{q-1}{q-z}\rfloor^{2}}
\end{split}
\end{eqnarray}
and
\begin{eqnarray}
\label{eq-comF-Y-C1-1}
\frac{F_{z}}{F_{M-Y}}&=\frac{\lfloor\frac{q-1}{q-z}\rfloor q^m}{(q-1) q^m+q^m}=\frac{\lfloor\frac{q-1}{q-z}\rfloor }{(q-1)+1}=\frac{\lfloor\frac{q-1}{q-z}\rfloor}{q}.
\end{eqnarray}
\begin{remark}
\label{re-2}
From \eqref{eq-comR-Y-C1-1} and \eqref{eq-comF-Y-C1-1}, when $\lambda \lfloor\frac{q-1}{q-z}\rfloor^{2}>1$, $R_z$ and $F_z$ are at least $\frac{1}{\lambda \lfloor\frac{q-1}{q-z}\rfloor^{2}}$ and $\frac{\lfloor\frac{q-1}{q-z}\rfloor}{q}$ times smaller than $R_{M-Y}$ and $F_{M-Y}$ respectively.
\end{remark}
\begin{example}
\label{ex-com1}
When $q=20$ and $\lambda=0.5$, the following table can be obtained by \eqref{eq-comR-Y-C1-1} and \eqref{eq-comF-Y-C1-1}.
\begin{table}[H]
  \centering
  \caption{Comparisons of the performances between scheme from \cite{YCTC} and Theorem \ref{th-special case 1}} \label{tab_compare th3}
  \normalsize{
  \begin{tabular}{|c|c|c|}
\hline
$z$&  $\frac{R_{z}}{R_{M-S}}<$    &$\frac{F_{z}}{F_{M-S}}$   \\ \hline
11 & 0.5 & 0.1\\ \hline
12 & 0.5 & 0.1\\ \hline
13 & 0.5 & 0.1\\ \hline
14 &0.222222222222222  &0.15\\ \hline
15 &0.222222222222222  &0.15\\ \hline
16 &0.125 & 0.2\\ \hline
17 &0.0555555555555556 & 0.3\\ \hline
18 &0.0246913580246914 &0.45\\ \hline
\end{tabular}}
\end{table}
\end{example}

If there is no integer $z$ satisfying $\frac{M}{N}=\frac{z}{q}$, a scheme in \eqref{eq-M-MNRF} based on the schemes in \eqref{eq-Sep-1} has the parameters
\begin{eqnarray}
\label{eq-MN-M-Sep-1}
\begin{split}
\frac{M}{N}&=\lambda'\frac{z}{q}+(1-\lambda')\frac{z+1}{q}\\
R_{M-C_2}&=\lambda' \frac{q-z}{\lfloor\frac{q-1}{q-z}\rfloor}+(1-\lambda') \frac{q-z-1}{\lfloor\frac{q-1}{q-z-1}\rfloor}< \frac{q-z}{\lfloor\frac{q-1}{q-z}\rfloor}\\
F_{M-C_2}&=\lfloor\frac{q-1}{q-z}\rfloor q^m+\lfloor\frac{q-1}{q-z-1}\rfloor q^m
\end{split}
\end{eqnarray}
for some positive integer $z$ with $\frac{z}{q}<\frac{M}{N}<\frac{z+1}{q}$ and a real number $0<\lambda< 1$. Similar to \eqref{eq-comR-Y-C1-1} and \eqref{eq-comF-Y-C1-1}, we have
\begin{eqnarray}
\label{eq-comR-Y-C1-2}
\frac{R_{M-C_2}}{R_{M-S}}&<\frac{\frac{q-z}{\lfloor\frac{q-1}{q-z}\rfloor}}{\lambda (q-1)+(1-\lambda) \frac{1}{q-1}}<\frac{1}{\lambda \lfloor\frac{q-1}{q-z}\rfloor^{2}}
\end{eqnarray}
and
\begin{eqnarray}
\label{eq-comF-Y-C1-2}
\begin{split}
\frac{F_{M-C_2}}{F_{M-S}}&=\frac{\lfloor\frac{q-1}{q-z}\rfloor q^m+\lfloor\frac{q-1}{q-z-1}\rfloor^t q^m}{(q-1) q^m+q^m}=\frac{\lfloor\frac{q-1}{q-z}\rfloor +\lfloor\frac{q-1}{q-z-1}\rfloor }{(q-1)+1}=\frac{\lfloor\frac{q-1}{q-z}\rfloor}{(q-1)+1} +\frac{\lfloor\frac{q-1}{q-z-1}\rfloor }{(q-1)+1}\\
&<\frac{1}{q-z}+\frac{1}{q-z-1}.
\end{split}
\end{eqnarray}
Similar to the Remark \ref{re-2}, we have that when $\lambda \lfloor\frac{q-1}{q-z}\rfloor^{2}>1$ and $q-z>1$, $R_{M-C_2}$ and $F_{M-C_2}$ are at least $\frac{1}{\lambda \lfloor\frac{q-1}{q-z}\rfloor^{2}}$ and $\frac{1}{q-z}+\frac{1}{q-z-1}$ times smaller than $R_{M-Y}$ and $F_{M-Y}$ respectively.

\section{Explicit Constructions}
\label{se-generalized}
In this section, four classes of PDAs in Table \ref{tab-main} are constructed.
\subsection{The first construction}
First let us consider an example. When $m=2$, $t=1$ and $q=3$, from the construction in \cite{SZG}, the following $(6,9,3,9)$ PDA can be obtained.
\begin{small}
\begin{eqnarray}
\begin{small}
\label{eq_PDA_3,2}
\mathbf{P}=\left(
  \begin{array}{ccc|ccc}
 *  & 11 & 3  & *  & 13 & 7\\
 1  & *  & 12 & *  & 14 & 8\\
 10  & 2  & *  & *  & 15 & 9\\
 *  & 14 & 6  & 1  & *  & 16\\
 4  & *  & 15 & 2  & *  & 17\\
 13 & 5  & *  & 3  & *  & 18\\
 *  & 17 & 9  & 10  & 4  & *\\
 7  & *  & 18 & 11 & 5  & *\\
 16 & 8  & *  & 12 & 6  & *
  \end{array}
\right)
\end{small}
\end{eqnarray}
\end{small}For the detailed constructions, the interested readers can be referred to \cite{SZG}. Replacing three integer entries of each column of $\mathbf{P}$ in \eqref{eq_PDA_3,2} by ``$*$"s elaborately, we have $\mathbf{P}_1$ in the following.
\begin{eqnarray*}
\mathbf{P}_1=\left(
  \begin{array}{ccc|ccc}
 * & * & 3 & * & * & 7\\
 1 & * & * & * & * & 8\\
 * & 2 & * & * & * & 9\\
 * & * & 6 & 1 & * & *\\
 4 & * & * & 2 & * & *\\
 * & 5 & * & 3 & * & *\\
 * & * & 9 & * & 4 & *\\
 7 & * & * & * & 5 & *\\
 * & 8 & * & * & 6 & *
\end{array}
\right)
\end{eqnarray*}
Then changing the integer in each entry, the following array $\mathbf{P}_2$ can be obtained.
\begin{eqnarray*}
\mathbf{P}_2=\left(
  \begin{array}{ccc|ccc}
 * & * & 2 & * & * & 4\\
 3 & * & * & * & * & 5\\
 * & 1 & * & * & * & 6\\
 * & * & 5 & 7 & * & *\\
 6 & * & * & 8 & * & *\\
 * & 4 & * & 9 & * & *\\
 * & * & 8 & * & 1 & *\\
 9 & * & * & * & 2 & *\\
 * & 7 & * & * & 3 & *
\end{array}
\right)
\end{eqnarray*}
It is interesting that the following array $\mathbf{P}'={\mathbf{P}_1\choose\mathbf{P}_2}$ is a $(6,18,12,9)$ PDA too.
\begin{small}
\begin{eqnarray}
\label{eq_PDA_3,2-2b}
\mathbf{P}'=\left(
  \begin{array}{ccc|ccc}
 * & * & 3 & * & * & 7\\
 1 & * & * & * & * & 8\\
 * & 2 & * & * & * & 9\\
 * & * & 6 & 1 & * & *\\
 4 & * & * & 2 & * & *\\
 * & 5 & * & 3 & * & *\\
 * & * & 9 & * & 4 & *\\
 7 & * & * & * & 5 & *\\
 * & 8 & * & * & 6 & *\\ \hline
 * & * & 2 & * & * & 4\\
 3 & * & * & * & * & 5\\
 * & 1 & * & * & * & 6\\
 * & * & 5 & 7 & * & *\\
 6 & * & * & 8 & * & *\\
 * & 4 & * & 9 & * & *\\
 * & * & 8 & * & 1 & *\\
 9 & * & * & * & 2 & *\\
 * & 7 & * & * & 3 & *
\end{array}
\right)
\end{eqnarray}
\end{small}Inspired by replacing method in the above example, we can propose our first construction. For ease of understanding, we will use $q$-ary sequences to represent the parameters $K$, $F$, $Z$ and $S$ of a PDA unless otherwise stated.
\begin{construction}
\label{con-general-1}
For any positive integers $q$, $z$, $m$ and $t$ with $0<z<q$ and $0<t<m$, let
\begin{eqnarray}
\label{eq-FK}
\begin{split}
\mathcal{F}&=\{{\bf a}=(a_0,a_1,\ldots,a_{m-1},\varepsilon_0,\ldots,\varepsilon_{t-1})\ |\ a_0, \ldots,a_{m-1}\in \mathbb{Z}_q,\varepsilon_0,\ldots,\varepsilon_{t-1}\in [0,\lfloor\textstyle\frac{q-1}{q-z}\rfloor)\},\\
\mathcal{K}&=\{{\bf b}=(b_0,b_1,\ldots,b_{t-1},\delta_0,\delta_1,\ldots,\delta_{t-1})\ |\
b_{0},\ldots,b_{t-1}\in\mathbb{Z}_q,0\leq \delta_0<\ldots<\delta_{t-1}<m,\}.
\end{split}
\end{eqnarray}
Then a $\lfloor\frac{q-1}{q-z}\rfloor^t q^{m}\times {m\choose t}q^t$ array $\mathbf{P}=(p_{{\bf a},{\bf b}})$ can be defined in the following way
\begin{eqnarray}\label{Eqn_Gen._PDA1}
p_{{\bf a}, {\bf b}}=\left\{
\begin{array}{ll}
(a_0, a_1,\ldots, b_{i}-\varepsilon_i(q-z),\ldots, a_{m-1},a_{\delta_0}-b_{0}-1,\ldots,a_{\delta_{t-1}}-b_{t-1}-1)_q & \textrm{if}~a_{\delta_i}\not\in X_{b_i,z}, \forall i\in [0,t)\\
* & \textrm{otherwise} \end{array}
\right.
\end{eqnarray}
where $X_{b_i,z}=\{b_i,b_i-1,\ldots, b_i-(z-1)\}$. Here the operations are performed modulo $q$.
\end{construction}
Based on the array generated by Construction \ref{con-general-1}, the proof of Theorem \ref{th-general case 1} can be obtained and is included in Appendix A.
\begin{example}
\label{ex_gen1}
Assume that $m=2$, $q=3$ and $t=1$. When $z=2$, $\lfloor\frac{q-1}{q-z}\rfloor=2$ holds. Then \eqref{eq-FK} can be written as follows.
$$\mathcal{F}=\{(a_0,a_1,\varepsilon)\ |\ a_0,a_1\in \mathbb{Z}_3,\varepsilon\in [0,1]\}\ \ \ \hbox{and}\ \ \ \ \mathcal{K}=\{(b_0 ,\delta_0 )\ |\ b_0 \in\mathbb{Z}_3,
0\leq \delta_0 \leq 1\}$$
By \eqref{Eqn_Gen._PDA1}, the following array can be obtained.
\begin{eqnarray}
\label{eq_tableR-32}
\begin{small}
\setlength{\arraycolsep}{0.25pt}
\begin{array}{c|cccccc}
(a_0,a_1,\varepsilon) \backslash (b_0 ,\delta_0 ) &
          (0,0)& (1,0) & (2,0) & (0,1) & (1,1) & (2,1) \\ \hline
(0,0,0)&  *    &*      &(2,0,0)&*      &*      &(0,2,0)\\[-0.1cm]
(1,0,0)&(0,0,0)&*      &*      &*      & *     &(1,2,0)\\[-0.1cm]
(2,0,0)&  *    &(1,0,0)&*      &*      &*      &(2,2,0) \\[-0.1cm]
(0,1,0)&*      &*      &(2,1,0)&(0,0,0)&*      &*\\[-0.1cm]
(1,1,0)&(0,1,0)&*      &*      &(1,0,0)&*      &*\\[-0.1cm]
(2,1,0)& *     &(1,1,0)&*      &(2,0,0)&*      &*\\[-0.1cm]
(0,2,0)& *     &*      &(2,2,0)&*      &(0,1,0)&* \\[-0.1cm]
(1,2,0)&(0,2,0)&*      &*      &*      &(1,1,0)&*\\[-0.1cm]
(2,2,0)& *     &(1,2,0)&*      &*      &(2,1,0)&*\\[-0.1cm]
(0,0,1)& *     &*      &(1,0,0)&*      &*      &(0,1,0)\\[-0.1cm]
(1,0,1)&(2,0,0)&*      &*      &*      &*      &(1,1,0)\\[-0.1cm]
(2,0,1)& *     &(0,0,0)&*      &*      &*      &(2,1,0)\\[-0.1cm]
(0,1,1)& *     &*      &(1,1,0)&(0,2,0)&*      &*\\[-0.1cm]
(1,1,1)&(2,1,0)&*      &*      &(1,2,0)&*      &*\\[-0.1cm]
(2,1,1)& *     &(0,1,0)&*      &(2,2,0)&*      &*\\[-0.1cm]
(0,2,1)& *     &*      &(1,2,0)&*      &(0,0,0)&*\\[-0.1cm]
(1,2,1)&(2,2,0)&*      &*      & *     &(1,0,0)&*\\[-0.1cm]
(2,2,1)& *     &(0,2,0)&*      &*      &(2,0,0)&*\\ \hline
\end{array}
\end{small}
\end{eqnarray}
In fact the array in \eqref{eq_tableR-32} is exactly the PDA in \eqref{eq_PDA_3,2-2b} when we replace the vectors by the integers in $[1,9]$.
\end{example}

\subsection{The second construction}
\label{sec_sepcial}
Let us also consider an example. When $m=2$ and $q=3$, from the construction in \cite{YCTC}, the following $(9,9,3,9)$ PDA can be obtained.
\begin{small}
\begin{eqnarray}
\label{eq-Y99}
\begin{small}
\mathbf{P}=\left(
  \begin{array}{ccc|ccc|ccc}
 *  & 11 & 3  & *  & 13 & 7  & *  & 1  & 10\\
 1  & *  & 12 & *  & 14 & 8  & 11 & *  & 2\\
 10  & 2  & *  & *  & 15 & 9  & 3  & 12 & *\\
 *  & 14 & 6  & 1  & *  & 16 & 13 & *  & 4\\
 4  & *  & 15 & 2  & *  & 17 & 5  & 14 & *\\
 13 & 5  & *  & 3  & *  & 18 & *  & 6  & 15\\
 *  & 17 & 9  & 10  & 4  & *  & 7  & 16 & *\\
 7  & *  & 18 & 11 & 5  & *  & *  & 8  & 17\\
 16 & 8  & *  & 12 & 6  & *  & 18 & *  & 9
  \end{array}
\right)
\end{small}
\end{eqnarray}
\end{small}For the detailed constructions, the interested readers can be referred to \cite{YCTC}. Similar to generating the array in \eqref{eq_PDA_3,2-2b}, the following $(9,18,12,9)$ PDA can be obtained.
\begin{small}
\begin{eqnarray}
\label{eq_PDA_918}
\mathbf{P}'
=\left(
  \begin{array}{ccc|ccc|ccc}
 * & * & 3 & * & * & 7 & * & 1 & *\\
 1 & * & * & * & * & 8 & * & * & 2\\
 * & 2 & * & * & * & 9 & 3 & * & *\\
 * & * & 6 & 1 & * & * & * & * & 4\\
 4 & * & * & 2 & * & * & 5 & * & *\\
 * & 5 & * & 3 & * & * & * & 6 & *\\
 * & * & 9 & * & 4 & * & 7 & * & *\\
 7 & * & * & * & 5 & * & * & 8 & *\\
 * & 8 & * & * & 6 & * & * & * & 9\\ \hline
 * & * & 2 & * & * & 4 & 1 & * & *\\
 3 & * & * & * & * & 5 & * & 2 & *\\
 * & 1 & * & * & * & 6 & * & * & 3\\
 * & * & 5 & 7 & * & * & * & 4 & *\\
 6 & * & * & 8 & * & * & * & * & 5\\
 * & 4 & * & 9 & * & * & 6 & * & *\\
 * & * & 8 & * & 1 & * & * & * & 7\\
 9 & * & * & * & 2 & * & 8 & * & *\\
 * & 7 & * & * & 3 & * & * & 9 & *
\end{array}
\right)
\end{eqnarray}
\end{small}It is worth noting that the replacing method in \eqref{eq_PDA_918} is more complex than that in \eqref{eq_PDA_3,2-2b} since the number of columns of the array in \eqref{eq-Y99} is grater than that of the array in \eqref{eq_PDA_3,2}. However we can still propose our second construction for any positive integers $m$, $q$ and $z$.
\begin{construction}
\label{con-special-1}
When $t=1$, \eqref{eq-FK} can be written as
\begin{eqnarray*}
\mathcal{F}=\{(a_0,a_1,\ldots,a_{m-1},\varepsilon)\ |\ a_0, \ldots,a_{m-1}\in \mathbb{Z}_q,\varepsilon\in [0,\lfloor{\textstyle\frac{q-1}{q-z}}\rfloor)\}\ \ \hbox{and}\ \ \ \mathcal{K}=\{(b,\delta)\ |\ b\in\mathbb{Z}_q,
0\leq \delta< m\}
\end{eqnarray*}for any positive integers $q$, $z$ and $m$ with $z<q$.
Let $\mathcal{K}_1=\{(b,m)\ |\ b\in\mathbb{Z}_q\}$. Define an array $\mathbf{H}=(\mathbf{P},\mathbf{C})$ where
\begin{itemize}
\item $\mathbf{P}=(p_{{\bf a}, {\bf b}})$, ${\bf a}\in \mathcal{F}, {\bf b}\in \mathcal{K}$, is the $\lfloor\frac{q-1}{q-z}\rfloor q^{m}\times mq$ array generated by
Construction \ref{con-general-1}.
\item  $\mathbf{C}=(c_{{\bf a},{\bf b}})$, ${\bf a}\in \mathcal{F}, {\bf b}\in \mathcal{K}_1$ is a $\lfloor\frac{q-1}{q-z}\rfloor q^{m}\times q$ array define in the following way.
\begin{eqnarray}\label{Eqn_Spec._PDA1}
c_{{\bf a},{\bf b}}=\left\{
\begin{array}{ll}
(a_0,\cdots,a_{m-1},b -(\sum_{l=0}^{m-1}a_l-\varepsilon(q-z))-1)_q & \textrm{if}~\sum_{l=0}^{m-1}a_l-\varepsilon(q-z)\not\in Y_{b ,z}\\
* & \textrm{otherwise}
\end{array}
\right.
\end{eqnarray}
where $Y_{b ,z}=\{b ,b +1,\ldots, b +(z-1)\}$. All the above operations are performed modulo $q$.
\end{itemize}
\end{construction}

Based on the array generated by Construction \ref{con-special-1}, the proof of Theorem \ref{th-special case 1} can be obtained and is included in Appendix B.
\begin{example}
\label{ex_special}
Suppose that $m=2$, $q=3$ and $z=2$. Then we have $\mathcal{K}_1=\{(b,2)\ |  b\in\mathbb{Z}_3\}$. By \eqref{Eqn_Gen._PDA1} and \eqref{Eqn_Spec._PDA1}, the following array can be obtained.
\begin{eqnarray}
\label{eq_tableR1-32Sum}
\setlength{\arraycolsep}{0.25pt}
\begin{small}
\begin{array}{c|ccccccccc}
(a_0,a_1,\varepsilon) \backslash (b,\delta)
       &(0,0)  &(1,0)  &(2,0)  &(0,1)  &(1,1)  &(2,1)  &(0,2)  &(1,2)  &(2,2)\\ \hline
(0,0,0)&*      &*      &(2,0,0)&*      &*      &(0,2,0)&*      &(0,0,0)&* \\[-0.1cm]
(1,0,0)&(0,0,0)&*      &*      &*      &*      &(1,2,0)&*      &*      &(1,0,0) \\[-0.1cm]
(2,0,0)&*      &(1,0,0)&*      &*      &*      &(2,2,0)&(2,0,0)&*      &* \\[-0.1cm]
(0,1,0)&*      &*      &(2,1,0)&(0,0,0)&*      &*      &*      &*      &(0,1,0) \\[-0.1cm]
(1,1,0)&(0,1,0)&*      &*      &(1,0,0)&*      &*      &(1,1,0)&*      &* \\[-0.1cm]
(2,1,0)&*      &(1,1,0)&*      &(2,0,0)&*      &*      &*      &(2,1,0)&* \\[-0.1cm]
(0,2,0)&*      &*      &(2,2,0)&*      &(0,1,0)&*      &(0,2,0)&*      &* \\[-0.1cm]
(1,2,0)&(0,2,0)&*      &*      &*      &(1,1,0)&*      &*      &(1,2,0)&* \\[-0.1cm]
(2,2,0)&*      &(1,2,0)&*      &*      &(2,1,0)&*      &*      &*      &(2,2,0) \\[-0.1cm]
(0,0,1)&*      &*      &(1,0,0)&*      &*      &(0,1,0)&(0,0,0)&*      &* \\[-0.1cm]
(1,0,1)&(2,0,0)&*      &*      &*      &*      &(1,1,0)&*      &(1,0,0)&* \\[-0.1cm]
(2,0,1)&*      &(0,0,0)&*      &*      &*      &(2,1,0)&*      &*      &(2,0,0) \\[-0.1cm]
(0,1,1)&*      &*      &(1,1,0)&(0,2,0)&*      &*      &*      &(0,1,0)&* \\[-0.1cm]
(1,1,1)&(2,1,0)&*      &*      &(1,2,0)&*      &*      &*      &*      &(1,1,0) \\[-0.1cm]
(2,1,1)&*      &(0,1,0)&*      &(2,2,0)&*      &*      &(2,1,0)&*      &* \\[-0.1cm]
(0,2,1)&*      &*      &(1,2,0)&*      &(0,0,0)&*      &*      &*      &(0,2,0) \\[-0.1cm]
(1,2,1)&(2,2,0)&*      &*      &*      &(1,0,0)&*      &(1,2,0)&*      &* \\[-0.1cm]
(2,2,1)&*      &(0,2,0)&*      &*      &(2,0,0)&*      &*      &(2,2,0)&* \\ \hline
\end{array}
\end{small}
\end{eqnarray}
\end{example}
The array in \eqref{eq_tableR1-32Sum} is exactly the PDA in \eqref{eq_PDA_918} when we replace the vectors by the integers in $[1,9]$.

%
%
%
%
%
%
\subsection{The first extended construction}
It is interesting that we can obtain the following $(12,9,6,9)$ PDA $\mathbf{P}''$ by copying or replacing three integer entries of each column of $\mathbf{P}$ in \eqref{eq_PDA_3,2} by ``$*$"s elaborately.
\begin{small}
\begin{eqnarray}
\label{eq_PDA_129}
\mathbf{P}''=\left(
  \begin{array}{ccc|ccc|ccc|ccc}
 * & * & 3 & * & * & 2 & * & * & 7 & * & * & 4 \\
 1 & * & * & 3 & * & * & * & * & 8 & * & * & 5\\
 * & 2 & * & * & 1 & * & * & * & 9 & * & * & 6\\
 * & * & 6 & * & * & 5 & 1 & * & * & 7 & * & *\\
 4 & * & * & 6 & * & * & 2 & * & * & 8 & * & *\\
 * & 5 & * & * & 4 & * & 3 & * & * & 9 & * & *\\
 * & * & 9 & * & * & 8 & * & 4 & * & * & 1 & *\\
 7 & * & * & 9 & * & * & * & 5 & * & * & 2 & *\\
 * & 8 & * & * & 7 & * & * & 6 & * & * & 3 & *
  \end{array}
\right)
\end{eqnarray}
\end{small}
By modifying \eqref{eq-FK}, the following construction can be obtained.
\begin{construction}
\label{cons-3}
For any positive integers $q$, $z$, $m$ and $t$ with $z<q$ and $t<m$, let
\begin{eqnarray}
\label{eq-F}
\mathcal{F}=\{(a_0,a_1,\ldots,a_{m-1})\ |\ a_0, \ldots,a_{m-1}\in \mathbb{Z}_q\}.
\end{eqnarray}
and
\begin{eqnarray}
\begin{split}
\label{eq-K}
\mathcal{K}=\{(b_0,b_1,\ldots,b_{t-1},\delta_0,\delta_1,\ldots,\delta_{t-1},\varepsilon_0,\ldots,\varepsilon_{t-1})\ |&
b_{0}, \ldots,b_{t-1}\in\mathbb{Z}_q,\\
&0\leq \delta_0<\ldots<\delta_{t-1}<m,
\varepsilon_0,\ldots,\varepsilon_{t-1}\in [0,\lfloor{\textstyle\frac{q-1}{q-z}}\rfloor)\}
\end{split}
\end{eqnarray}
Using the same rule defined in \eqref{Eqn_Gen._PDA1}, we can obtain a $q^{m}\times {m\choose t}\lfloor\frac{q-1}{q-z}\rfloor^t q^t$ array.
\end{construction}
\begin{example}
\label{ex_gen1}
When $m=2$, $q=3$ and $z=2$, \eqref{eq-F} and \eqref{eq-K} can be written as
$$\mathcal{F}=\{(a_0,a_1)\ |\ a_0,a_1\in \mathbb{Z}_3\}\ \ \ \hbox{and}\ \ \ \ \mathcal{K}=\{(b,\delta,\varepsilon)\ |
b\in\mathbb{Z}_3,0\leq \delta\leq 1,\varepsilon\in [0,1]\}.$$
By \eqref{Eqn_Gen._PDA1}, the following array can be obtained. It is easy to check that this array is a $(12,9,6,9)$ PDA.
\begin{small}
\begin{eqnarray}
\setlength{\arraycolsep}{0.25pt}
\label{eq_tableC-31}
\begin{array}{c|cccccccccccc}
(a_0,a_1) \backslash (b,\delta,\varepsilon)
     &(0,0,0)&(1,0,0)&(2,0,0)&(0,0,1)&(1,0,1)&(2,0,1)&(0,1,0)&(1,1,0)&(2,1,0)&(0,1,1)&(1,1,1) &(2,1,1) \\ \hline
(0,0)& *     &*      &(2,0,0)& *     &*      &(1,0,0)        &*      &*      &(0,2,0)&*      &*      &(0,1,0)\\[-0.1cm]
(1,0)&(0,0,0)& *     &*      &(2,0,0)& *     &*              &*      &*      &(1,2,0)&*      &*      &(1,1,0)\\[-0.1cm]
(2,0)& *     &(1,0,0)&*      &*      &(0,0,0)&*              & *     &*      &(2,2,0)& *     &*      &(2,1,0)\\[-0.1cm]
(0,1)&*      &*      &(2,1,0)&*      &*      &(1,1,0)        &(0,0,0)&*      &*      &(0,2,0)&*      &   *   \\[-0.1cm]
(1,1)&(0,1,0)&  *    &  *    &(2,1,0)&  *    &*              &(1,0,0)&*      &*      &(1,2,0)&*      &   *   \\[-0.1cm]
(2,1)&*      &(1,1,0)&*      &*      &(0,1,0)&*              &(2,0,0)&*      &*      &(2,2,0)&*      &   *   \\[-0.1cm]
(0,2)&*      &*      &(2,2,0)&*      &*      &(1,2,0)        &*      &(0,1,0)&   *   &*      &(0,0,0)&   *   \\[-0.1cm]
(1,2)&(0,2,0)&*      &*      &(2,2,0)&*      &*              &*      &(1,1,0)&   *   &*      &(1,0,0)&   *   \\[-0.1cm]
(2,2)&*      &(1,2,0)&*      &*      &(0,2,0)&*              &*      &(2,1,0)&   *   &*      &(2,0,0)&   *   \\ \hline
\end{array}
\end{eqnarray}
\end{small}
\end{example}
The array in \eqref{eq_tableC-31} is exactly the PDA in \eqref{eq_PDA_129} when we replace the vectors by the integers in $[1,9]$. Similar to the proof of Theorem \ref{th-general case 1}, Theorem \ref{th-general case 2} can be proved.

\subsection{The second extended construction}
Similar to generating the PDA in \eqref{eq_PDA_129}, we can obtain the following $(15,9,6,9)$ PDA $\mathbf{P}''$ by copying or replacing three integer entries of each column of $\mathbf{P}$ in \eqref{eq-Y99} by ``$*$"s elaborately.

\begin{small}
\begin{eqnarray}
\label{eq_PDA_159}
\mathbf{P}''=\left(
  \begin{array}{ccc|ccc|ccc|ccc|ccc}
 * & * & 3 & * & * & 2 & * & * & 7 & * & * & 4 & * & 1 & *\\
 1 & * & * & 3 & * & * & * & * & 8 & * & * & 5 & * & * & 2\\
 * & 2 & * & * & 1 & * & * & * & 9 & * & * & 6 & 3 & * & *\\
 * & * & 6 & * & * & 5 & 1 & * & * & 7 & * & * & * & * & 4\\
 4 & * & * & 6 & * & * & 2 & * & * & 8 & * & * & 5 & * & *\\
 * & 5 & * & * & 4 & * & 3 & * & * & 9 & * & * & * & 6 & *\\
 * & * & 9 & * & * & 8 & * & 4 & * & * & 1 & * & 7 & * & *\\
 7 & * & * & 9 & * & * & * & 5 & * & * & 2 & * & * & 8 & *\\
 * & 8 & * & * & 7 & * & * & 6 & * & * & 3 & * & * & * & 9
  \end{array}
\right)
\end{eqnarray}
\end{small}
\begin{construction}
\label{con-special-2}
When $t=1$, \eqref{eq-F} and \eqref{eq-K} can be written as
\begin{eqnarray*}
\mathcal{F}=\{(a_0,a_1,\ldots,a_{m-1})\ |\ a_0, \ldots,a_{m-1}\in \mathbb{Z}_q\}\ \ \hbox{and}\ \ \ \mathcal{K}=\{(b,\delta,\varepsilon)\ |\
b\in\mathbb{Z}_q,0\leq \delta\leq m,\varepsilon\in [0,\lfloor{\textstyle\frac{q-1}{q-z}}\rfloor)\}
\end{eqnarray*}
for any positive integers $q$, $z$ and $m$ with $z<q$.
\begin{itemize}
\item Using the same rule defined in \eqref{Eqn_Gen._PDA1}, we can obtain a $q^{m}\times m\lfloor\frac{q-1}{q-z}\rfloor q$ array $\mathbf{P}=(p_{{\bf a},{\bf b}})$, ${\bf a}\in \mathcal{F}, {\bf b}\in \mathcal{K}$.
\item Let $\mathcal{K}_1=\{(b,m)\ |\ b\in\mathbb{Z}_q\}$. We can define a $q^{m}\times q$ array $\mathbf{C}=(c_{{\bf a},{\bf b}})$, ${\bf a}\in \mathcal{F}, {\bf b}\in \mathcal{K}_1$ in the following way.
\begin{eqnarray}\label{Eqn_Spec._PDA2}
c_{{\bf a},{\bf b}}=\left\{
\begin{array}{ll}
(a_0,\cdots,a_{m-1},b-\sum_{l=0}^{m-1}a_l-1)_q& \textrm{if}~\sum_{l=0}^{m-1}a_l\not\in Y_{b,z}\\
*  & \textrm{otherwise}
\end{array}
\right.
\end{eqnarray}
where $Y_{b,z}=\{b,b+1,\ldots, b+(z-1)\}$.  All the above operations are performed modulo $q$.
\end{itemize}
\end{construction}
Based on the array generated by Construction \ref{con-special-2}, the proof of Theorem \ref{th-special case 2} can be obtained and is included in Appendix C.
\section{Conclusion}
\label{conclusion}
In this paper, four classes of new schemes were constructed. Our new constructions did not only include some previously known constructions as special cases, but also had the following advantages: In the first two classes, $F$ growed sub-exponentially with $K$; And the other two classes could significantly reduce the value of $F$ by just increasing some fixed times of $R$ comparing with MN PDAs. In addition, an exact tradeoff between $R$ and $F$ was proposed.

\section*{Appendix A:Proof of Theorem \ref{th-general case 1}}
\begin{proof}
It is easy to check that there are exactly $q^{m-t}{(q-z)^t}\lfloor\frac{q-1}{q-z}\rfloor^t$ vector entries in each column. This implies that $Z=(q^m-q^{m-t}{(q-z)^t})\lfloor\frac{q-1}{q-z}\rfloor^t$. So C$1$ holds.
From \eqref{Eqn_Gen._PDA1}, the vector set of $\mathbf{P}$ in Construction \ref{con-general-1} is
$$\mathcal{S}=\{(a_0,a_1,\ldots,a_{m-1},a_{m},\ldots,a_{m+t-1})\ |\ a_0, \ldots,a_{m-1}\in \mathbb{Z}_q, a_{m+1},\ldots,a_{m+t-t}\in[0,q-z)\}.$$
So $S=|\mathcal{S}|=q^m (q-z)^t$. Each ${\bf s}=(s_0,s_1,\ldots,s_{m+t-1})\in \mathcal{S}$ appears in the entries at row ${\bf a}$ and column ${\bf b}$ of $\mathbf{P}$,
if and only if
\begin{eqnarray}
\begin{split}
\label{eq-Gen.eq3}
&{\bf a}=(a_0,\ldots,a_{\delta_i},\cdots, a_{m-1},\varepsilon_0,\ldots,\varepsilon_{t-1})=(s_0,\ldots,s_{\delta_i}+s_{m+i}+\varepsilon_{i}(q-z)+1,\cdots, s_{m-1},
\varepsilon_0,\ldots,\varepsilon_{t-1} )_q
\ \ \hbox{and}\\
&{\bf b}=(b_{0},b_{1},\ldots,b_{t-1},\delta_0,\delta_1,\ldots,\delta_{t-1})
=
(s_{\delta_0}+\varepsilon_{0}(q-z),s_{\delta_1}+\varepsilon_{1}(q-z),\ldots,
s_{\delta_{t-1}}+\varepsilon_{t-1}(q-z),\delta_0,\delta_1,\ldots,\delta_{t-1}).
\end{split}\end{eqnarray}

Clearly, for any fixed $\delta_0$, $\delta_1$, $\ldots$, $\delta_{t-1}$ and $\varepsilon_0$, $\ldots$, $\varepsilon_{t-1}$, there is exactly a unique pair of vectors ${\bf a}$ and ${\bf b}$, such that $p_{{\bf a},{\bf b}}={\bf s}$. Based on the above observation, C$2$ is clear.
Now, let us consider C$3$. Suppose there exists another two vectors ${\bf a}'$ and ${\bf b}'$ with $p_{{\bf a}',{\bf b}'}={\bf s}$. Then there must exist another $2t$ integers $
\delta'_0$, $\ldots$, $\delta'_{t-1}$ and $\varepsilon'_0$, $\ldots$, $\varepsilon'_{t-1}$ satisfying
\begin{eqnarray}
\begin{split}
\label{eq-Gen.eq4}
&{\bf a}'=(a'_0,\ldots,a'_{\delta'_i},\cdots,a'_{m-1},
\varepsilon'_0,\ldots,\varepsilon'_{t-1})=(s_0,\ldots,s_{\delta'_i}+s_{m+i}+\varepsilon'_{i}(q-z)+1,\cdots,s_{m-1},
\varepsilon'_0,\ldots,\varepsilon'_{t-1} )_q\ \ \hbox{and}\\
&{\bf b}'=(b'_{0},b'_{1},\ldots,b'_{t-1},\delta'_0,\delta'_1,\ldots,\delta'_{t-1})
 =
(s_{\delta'_0}+\varepsilon'_{0}(q-z),s_{\delta'_1}+\varepsilon'_{1}(q-z),\ldots,
s_{\delta'_{t-1}}+\varepsilon'_{t-1}(q-z),\delta'_0,\delta'_1,\ldots,\delta'_{t-1}).
\end{split}
\end{eqnarray}
It is sufficient to consider the following cases, where all the operations are performed modulo $q$.
\begin{itemize}
\item If $\{\delta_0$, $\ldots$, $\delta_{t-1}\}\neq\{\delta'_0$, $\ldots$, $\delta'_{t-1}\}$, there at least exist two integers $i,i'\in [0,t)$ such that $\delta_i\not\in \{\delta'_0$, $\ldots$, $\delta'_{t-1}\}$ and $\delta'_{i'}\not\in \{\delta_0$, $\ldots$, $\delta_{t-1}\}$. Now we show C$3$-a) and C$3$-b) hold respectively.
\begin{itemize}
\item If ${\bf s}$ occurs in a row at least twice, we have ${\bf a}={\bf a}'$. This implies that $a_{\delta_i}=a'_{\delta_i}$. From \eqref{eq-Gen.eq3} and \eqref{eq-Gen.eq4}, we have
\begin{eqnarray*}
\label{eq-Gen1-eq5}
a_{\delta_{i}}=s_{\delta_i}+s_{m+i}+\varepsilon_{i}(q-z)+1,\ \ \ \ \ \ \ \ \ \ a'_{\delta_i}=s_{\delta_i}.
\end{eqnarray*}
This implies that $s_{m+i}+\varepsilon_{i}(q-z)+1=0$. This is impossible since
$$1\leq s_{m+i}+\varepsilon_{i}(q-z)+1<q-z+(\frac{q-1}{q-z}-1)(q-z)+1=q$$
by the facts $s_{m+i}\in[0,q-z)$ and $\varepsilon\in [0,\lfloor\frac{q-1}{q-z}\rfloor)$. So C$3$-a) holds.
\item First from \eqref{eq-Gen.eq3} and \eqref{eq-Gen.eq4}, we have
\begin{eqnarray*}
\label{eq-Gen1-eq6}
a_{\delta'_{i'}}=s_{\delta'_{i'}},\ \ \ \ b'_{i'}=s_{\delta'_{i'}}+\varepsilon'_{i'}(q-z),\ \ \ \  a'_{\delta_i}=s_{\delta_i}, \ \ \ \ b'_{i}=s_{\delta_i}+\varepsilon_{i'}(q-z)
\end{eqnarray*}
If $p_{{\bf a},{\bf b}'}\neq *$, we have $a_{\delta'_{i'}}\in \{b'_{i'}+1,b'_{i'}+2,\ldots,b'_{i'}+(q-z)\}$. Then there exists an integer $j\in [1,q-z)$ such that $a_{\delta'_{i'}}=b'_{i'}+j$, i.e.,
$$a_{\delta'_{i'}}=s_{\delta'_{i'}}=s_{\delta'_{i'}}+\varepsilon'_{i'}(q-z)+j.$$
This implies that $\varepsilon'_{i'}(q-z)+j=0$. This is impossible since
$$1\leq \varepsilon'_{i'}(q-z)+j<(\frac{q-1}{q-z}-1)(q-z)+(q-z)=q-1$$
by the fact $\varepsilon'_{i'}\in [0,\lfloor\frac{q-1}{q-z}\rfloor)$. Similarly we can also show that $p_{{\bf a}',{\bf b}}=*$ too.
\end{itemize}
\item If $\{\delta_0$, $\ldots$, $\delta_{t-1}\}=\{\delta'_0$, $\ldots$, $\delta'_{t-1}\}$, we have $(\varepsilon_0,\ldots,\varepsilon_{t-1})\neq(\varepsilon'_0,\ldots,\varepsilon'_{t-1})$ by \eqref{eq-Gen.eq3} and \eqref{eq-Gen.eq4}. So there exist integer $i$ such that $\varepsilon_i\neq \varepsilon'_i$. Then we have
\begin{eqnarray}
\label{eq-Gen1-eq7}
b_{i}=s_{\delta_{i}}+\varepsilon_{i}(q-z),\ \ \ \  a_{\delta_{i}}=s_{m+i}+b_{i}+1, \ \ \ \ b'_{i}=s_{\delta_i}+\varepsilon'_{i}(q-z),\ \ \ \ a'_{\delta_i}=s_{m+i}+b'_{i}+1.
\end{eqnarray}
    Now we show C$3$-a) and C$3$-b) hold respectively.
    \begin{itemize}
\item If ${\bf s}$ occurs in a column at least twice, we have ${\bf b}={\bf b}'$. This implies that $b_{\delta_i}=b'_{\delta_i}$. By the first and third items in \eqref{eq-Gen1-eq7}, $\varepsilon_{i}(q-z)=\varepsilon'_{i}(q-z)$ holds. Clearly this is impossible since $\varepsilon_{i}, \varepsilon'_{i}\in [0,\lfloor\frac{q-1}{q-z}\rfloor)$.  So C$3$-a) holds.
\item If $p_{{\bf a},{\bf b}'}\neq *$, we have $a_{\delta_{i}}\in \{b'_{i}+1,b'_{i}+2,\ldots,b'_{i}+(q-z)\}$. Then there exists an integer $j\in [0,q-z)$ such that $a_{\delta_{i}}=b'_{i}+j$. By the first, second and third items in \eqref{eq-Gen1-eq7},
\begin{eqnarray}
\label{eq-Gen1-eq8}
s_{m+i}+s_{\delta_{i}}+\varepsilon_{i}(q-z)+1=s_{\delta_i}+\varepsilon'_{i}(q-z)+j.
\end{eqnarray}
If $\varepsilon_i<\varepsilon'_{i}$, we have $s_{m+i}+1=(\varepsilon'_{i}-\varepsilon_i)(q-z)+j$. This is impossible since
$$1\leq s_{m+i}+1\leq q-z\ \ \ \ \ \ \hbox{and}\ \ \ \ \ \ \ \ q-z+1\leq (\varepsilon'_{i}-\varepsilon_i)(q-z)+j<q$$
by the facts $0\leq \varepsilon_{i}$, $\varepsilon'_{i}<\lfloor\frac{q-1}{q-z}\rfloor$ and $1\leq j,s_{m+i}<q-z$. Clearly \eqref{eq-Gen1-eq8} does not hold either if $\varepsilon_i>\varepsilon'_{i}$. So we have $p_{{\bf a},{\bf b}'}= *$. Similarly we can also show that $p_{{\bf a}',{\bf b}}=*$ too.
\end{itemize}
\end{itemize}
\end{proof}

\section*{Appendix B:Proof of Theorem \ref{th-special case 1}}
\begin{proof}
From Construction \ref{con-special-1}, and the proof of Theorem \ref{th-general case 1}, we have that $\mathbf{P}$ is a $(mq,\lfloor\frac{q-1}{q-z}\rfloor q^{m},z\lfloor\frac{q-1}{q-z}\rfloor q^{m-1},(q-z)q^{m})$ PDA with
the vector set
$$\mathcal{S}=\{(s_0,s_1,\ldots,s_{m})\ |\  s_0,s_1,\ldots,s_{m-1}\in [0,q),\ s_m\in[0,q-z)\}.$$
From \eqref{Eqn_Spec._PDA1}, it is easy to count that each column of $\mathbf{C}$ has exactly $z\lfloor\frac{q-1}{q-z}\rfloor q^{m-1}$ stars and the vector set of $\mathbf{C}$ is $\mathcal{S}$. So C$1$ and C$2$ hold.
Now it is sufficient to verify C$3$. For any two distinct vector entries, say $h_{{\bf a},{\bf b}}$ and $h_{{\bf a}',{\bf b}'}$, assume that $h_{{\bf a},{\bf b}}=h_{{\bf a}',{\bf b}'}\in \mathcal{S}$ where
\begin{eqnarray*}
{\bf a}=(a_{0},a_1,\ldots,a_{m-1},\varepsilon)_q,\ \ \ \ {\bf a}'=(a'_{0},a'_1,\ldots,a'_{m-1},\varepsilon')_q,\ \ \ \ {\bf b}=(b,\delta), \ \ \ \  {\bf b}'=(b',\delta').
\end{eqnarray*}
We only need to consider the following cases, where all the operations are performed modulo $q$.

\begin{itemize}
\item When $\delta\in [0,m)$ and $\delta' =m$, we have $h_{{\bf a},{\bf b}}=p_{{\bf a},{\bf b}}$ and
$h_{{\bf a}',{\bf b}'}= c_{{\bf a}',{\bf b}'}$ in distinct columns. From \eqref{Eqn_Gen._PDA1} and \eqref{Eqn_Spec._PDA1} we have
\begin{eqnarray}
\label{eq-Spec1-eq1}
a_{\delta}-b -1=b' -(\sum_{l=0}^{m-1}a'_l-\varepsilon' (q-z))-1,\ \ \ a'_{\delta}=b -\varepsilon(q-z),\ \ \ \ a_l=a'_l
\end{eqnarray}
for any $l\in[0,m)\setminus\{\delta\}$. From the second equation in \eqref{eq-Spec1-eq1}, we can show $p_{{\bf a},{\bf b}'}$ and $c_{{\bf a}',{\bf b}}$ in distinct rows too. Otherwise we have $a_l=a'_l$ for any $l=0,1,\ldots,m-1$ and $\varepsilon=\varepsilon'$. Then combining the two equations in \eqref{eq-Spec1-eq1} we have
$$a_{\delta}-b =b' -(\sum_{l=0}^{m-1}a'_l-\varepsilon' (q-z))=-\varepsilon(q-z).$$
Since $\sum_{l=0}^{m-1}a'_l-\varepsilon' (q-z)\not\in Y_{b,z}$, let $\sum_{l=0}^{m-1}a'_l-\varepsilon' (q-z)=b'-j$ where $j\in [1,q-z]$. Then the above equation can be written as
$$b' -(\sum_{l=0}^{m-1}a'_l-\varepsilon' (q-z))=b'-(b'-j)=-\varepsilon(q-z).$$
This implies that $j+\varepsilon (q-z)=0$. This is impossible since
\begin{eqnarray}
\label{eq-Spec1-eq2}
1\leq j+\varepsilon(q-z)<(q-z)+(\frac{q-1}{q-z}-1)(q-z)=q-1
\end{eqnarray}
by the fact $\varepsilon\in [0,\lfloor\frac{q-1}{q-z}\rfloor)$. So C$3$-a) holds. Now let us consider the entries $c_{{\bf a},{\bf b}'}$ and  $p_{{\bf a}',{\bf b}}$ in the following cases.
\begin{itemize}
\item If $c_{{\bf a},{\bf b}'}$ is a vector, $\sum_{l=0}^{m-1}a_l-\varepsilon (q-z)\in \{b' -(q-z),b' -(q-z-1),\ldots, b' -1\}$. That is, there must exist an integer $j\in \{1,2,\ldots q-z\}$ such that $\sum_{l=0}^{m-1}a_l-\varepsilon (q-z)=b'-j$. From \eqref{eq-Spec1-eq1},
\begin{eqnarray*}
\sum_{l=0}^{m-1}a_l-\varepsilon (q-z)&=&\sum_{l\in [0,m)\setminus\{\delta\}}a_l+a_{\delta}-\varepsilon (q-z)\\
&=&\sum_{l\in [0,m)\setminus\{\delta\}}a_l+(b +b' -(\sum_{l=0}^{m-1}a'_l-\varepsilon' (q-z)))-\varepsilon (q-z)\\
&=&\sum_{l\in [0,m)\setminus\{\delta\}}a_l+(b +b' -\sum_{l\in [0,m)\setminus\{\delta\}}a'_l-a'_{\delta}+\varepsilon' (q-z))-\varepsilon (q-z)\\
&=&b +b' -a'_{\delta}+\varepsilon' (q-z)-\varepsilon (q-z)\\
&=&b +b' -(b -\varepsilon(q-z))+\varepsilon' (q-z)-\varepsilon (q-z)\\
&=&b' +\varepsilon' (q-z).
\end{eqnarray*}
So we have $b'-j=b' +\varepsilon' (q-z)$, i.e., $j+\varepsilon' (q-z)=0$. This is impossible by the similar proof in \eqref{eq-Spec1-eq2}. So $c_{{\bf a},{\bf b}'}=*$.
\item If $p_{{\bf a}',{\bf b}}$ is a vector, $a'_{\delta}\in \{b +1,b +2,\ldots,b +q-z\}$. There must exist an integer $j\in \{1,2,\ldots q-z\}$ such that $a'_{\delta}=b+j$. From \eqref{eq-Spec1-eq1}, we have $b +j=b -\varepsilon(q-z)$, i.e., $j+\varepsilon(q-z)=0$. This is impossible by \eqref{eq-Spec1-eq2}. So $p_{{\bf a}',{\bf b}}=*$.
\end{itemize}
So  C$3$-b) holds.
\item When $\delta=\delta' =m$, we have $h_{{\bf a},{\bf b}}=c_{{\bf a},{\bf b}}$ and  $h_{{\bf a}',{\bf b}'}=c_{{\bf a}',{\bf b}'}$. From \eqref{Eqn_Spec._PDA1} we have
\begin{eqnarray}\label{eq-Spec1-eq3}
b -(\sum_{l=0}^{m-1}a_l-\varepsilon(q-z))-1=b' -(\sum_{l=0}^{m-1}a'_l-\varepsilon' (q-z))-1,\ \ \ a_l=a'_l
\end{eqnarray}
for any $l\in[0,m)$. Clearly $\varepsilon\neq \varepsilon' $ always holds. Otherwise we have ${\bf a}={\bf a}'$ and $b =b' $, i.e., ${\bf b}={\bf b}'$. This is impossible. If ${\bf b}={\bf b}'$, we have $\varepsilon=\varepsilon'$ since $0\leq \varepsilon' (q-z), \varepsilon' (q-z)<q$ by the fact $\varepsilon$, $\varepsilon'\in  [0,\lfloor\frac{q-1}{q-z}\rfloor)$. So C$3$-a) is clear. Now let us consider C$3$-b). If $c_{{\bf a},{\bf b}'}$ is a vector, there must exist an integer $x\in \{1,2,\ldots q-z\}$ satisfying
$$\sum_{l=0}^{m-1}a_l-\varepsilon(q-z)=b' -x.$$

 Since $c_{{\bf a},{\bf b}}$ is a vector entry, there exist an integer $y\in \{1,2,\ldots q-z\}$ such that
 $$(\sum_{l=0}^{m-1}a_l-\varepsilon(q-z))=b -y.$$
 So we have
    \begin{eqnarray}\label{eqq-1}
    (\sum_{l=0}^{m-1}a_l-\varepsilon (q-z))=b'-x=b-y.
    \end{eqnarray}
Moreover, from \eqref{eq-Spec1-eq3} we have
\begin{eqnarray}\label{eqq-2}
b +\varepsilon(q-z)=b' +\varepsilon' (q-z).
\end{eqnarray}
Combining \eqref{eqq-1} and \eqref{eqq-2}, we have $x-y=(\varepsilon-\varepsilon' )(q-z)$. From the above discussion, we know that $\varepsilon\neq \varepsilon' $, i.e., $(\varepsilon-\varepsilon' )(q-z)\neq 0$. So we have $x\neq y$. Furthermore, $(q-z)|(x-y)$ holds since  $q-z\leq |(\varepsilon-\varepsilon' )(q-z)|< q-1$. Clearly this is impossible by the fact that $1\leq x\neq y\leq q-z$. So $c_{{\bf a},{\bf b}'}=*$. Similarly we can also show that $c_{{\bf a}',{\bf b}}=*$. So C$3$-b) holds.
\end{itemize}
\end{proof}

\section*{Appendix C: Proof of Theorem \ref{th-special case 2}}
\begin{proof}
Similar to the proof of Theorem \ref{th-special case 1}, denote $\mathbf{H}=(\mathbf{P},\mathbf{C})$. From Construction \ref{con-special-2} $\mathbf{P}$ is a $(m\lfloor\frac{q-1}{q-z}\rfloor q$, $q^m$, $zq^m$, $(q-z)q^{m})$ PDA with
the vector set
$$\mathcal{S}=\{(s_0,s_1,\ldots,s_{m})\ |\  s_0,s_1,\ldots,s_{m-1}\in [0,q),\ s_m\in[0,q-z)\}.$$
From \eqref{Eqn_Spec._PDA2}, it is easy to count that each column of $\mathbf{C}$ has exactly $zq^{m-1}$ stars and the vector set of $\mathbf{C}$ is $\mathcal{S}$. So C$1$ and C$2$ hold.
Now it is sufficient to verify C$3$. For any two distinct vector entries, say $h_{{\bf a},{\bf b}}$ and $h_{{\bf a}',{\bf b}'}$, assume that $h_{{\bf a},{\bf b}}=h_{{\bf a}',{\bf b}'}\in \mathcal{S}$ where
\begin{eqnarray*}
{\bf a}=(a_{0},a_1,\ldots,a_{m-1})_q,\ \ \ \ {\bf a}'=(a'_{0},a'_1,\ldots,a'_{m-1})_q,\ \ \ \ {\bf b}=(b,\delta,\varepsilon), \ \ \ \  {\bf b}'=(b',\delta',\varepsilon').
\end{eqnarray*}
We only need to consider the following cases, where all the operations are performed modulo $q$.
\begin{itemize}
\item When $\delta\in [0,m)$ and $\delta'=m$, we have $p_{{\bf a},{\bf b}}$ and  $c_{{\bf a}',{\bf b}'}$ in distinct columns. From \eqref{Eqn_Gen._PDA1} and \eqref{Eqn_Spec._PDA2} we have
\begin{eqnarray}\label{eq-Spec2-eq1}
a_{\delta}-b-1=b'-\sum_{l=0}^{m-1}a'_l-1,\ \ \ a'_{\delta}=s_{\delta}=b-\varepsilon (q-z),\ \ \ \ a_l=a'_l
\end{eqnarray}
for any $l\in[0,m)\setminus\{\delta\}$. From the second equation in \eqref{eq-Spec2-eq1}, we can show $p_{{\bf a},{\bf b}}$ and  $c_{{\bf a}',{\bf b}'}$ in distinct rows. Otherwise $a_{\delta}=a'_{\delta}$ holds. By \eqref{eq-Spec2-eq1} the following equation can be obtained.
$$a_{\delta}=a'_{\delta}=b+b'-\sum_{l=0}^{m-1}a'_l=b-\varepsilon(q-z)$$
Since $\sum_{l=0}^{m-1}a'_l\not\in Y_{b',z}$, there exists an integer $j\in\{1,2,\ldots,q-z\}$ such that $\sum_{l=0}^{m-1}a'_l=b'-j$. Combining the above formula we have
$$b+b'-(b'-j)=b-\varepsilon(q-z)$$
i.e., $j+\varepsilon(q-z)=0$. This is impossible since
\begin{eqnarray}
\label{eq-Spec2-eq2}
1\leq j+\varepsilon(q-z)<(q-z)+(\frac{q-1}{q-z}-1)(q-z)=q-1
\end{eqnarray}
by the fact $\varepsilon\in [0,\lfloor\frac{q-1}{q-z}\rfloor)$. So C$3$-a) holds. Now let us consider the entries $p_{{\bf a},{\bf b}'}$ and  $p_{{\bf a}',{\bf b}}$ in the following cases.
\begin{itemize}
\item If $c_{{\bf a},{\bf b}'}$ is a vector, we have $\sum_{l=0}^{m-1}a_l\in \{b' -(q-z),b' -(q-z-1),\ldots, b' -1\}$. There must exist an integer $j\in \{1,2,\ldots q-z\}$ such that $\sum_{l=0}^{m-1}a_l=b'-j$. From \eqref{eq-Spec2-eq1}, we have
\begin{eqnarray*}
\sum_{l=0}^{m-1}a_l&=&\sum_{l\in [0,m)\setminus\{\delta\}}a_l+a_{\delta}\\
&=&\sum_{l\in [0,m)\setminus\{\delta\}}a_l+(b+b'-\sum_{l=0}^{m-1}a'_l)\\
&=&\sum_{l\in [0,m)\setminus\{\delta\}}a_l+(b+b'-\sum_{l\in [0,m)\setminus\{\delta\}}a'_l-a'_{\delta})\\
&=&b+b'-a'_{\delta}\\
&=&b+b'-(b-\varepsilon (q-z))\\
&=&b'+\varepsilon (q-z)
\end{eqnarray*}
So we have $b'-j=b'+\varepsilon (q-z)$, i.e., $j+\varepsilon (q-z)=0$. This is impossible by \eqref{eq-Spec2-eq2}. So $c_{{\bf a},{\bf b}'}=*$.
\item If $p_{{\bf a}',{\bf b}}$ is a vector, we have $a'_{\delta}\in \{b+1,b+2,\ldots,b+q-z\}$. There must exist an integer $j\in \{1,2,\ldots q-z\}$ such that $a'_{\delta}=b+j$. From \eqref{eq-Spec2-eq1}, we have $b+j=b-\varepsilon (q-z)$, i.e., $j+\varepsilon(q-z)=0$. This is impossible by \eqref{eq-Spec2-eq2}. So $p_{{\bf a}',{\bf b}}=*$.
\end{itemize}
So C$3$-b) holds.
\item When $\delta=\delta'=m$, from \eqref{Eqn_Spec._PDA2} we have
\begin{eqnarray*}\label{eq-k'1=k''1=m}
b-\sum_{l=0}^{m-1}a_l-1=b'-\sum_{l=0}^{m-1}a'_l-1,\ \ \ a_l=a'_l
\end{eqnarray*}
for any $l\in[0,m)$. This implies that ${\bf a}={\bf a}'$ and ${\bf b}={\bf b}'$, a contradiction to our hypothesis. So C$3$-b) holds.
\end{itemize}
\end{proof}

\end{document}